\documentclass[twocolumn,showpacs,preprintnumbers,amsmath,amssymb]{revtex4}

\usepackage{graphicx}
\usepackage{dcolumn}
\usepackage{bm}

\newcommand{\bq}{\begin{equation}}
\newcommand{\eq}{\end{equation}}
\newcommand{\bqa}{\begin{eqnarray}}
\newcommand{\eqa}{\end{eqnarray}}
\newcommand{\nn}{\nonumber \\}

\def\be     {\begin{equation}}
\def\ee     {\end{equation}}
\def\bea        {\begin{eqnarray}}
\def\eea        {\end{eqnarray}}
\def\bnn    {\begin{eqnarray*}}
\def\enn    {\end{eqnarray*}}

\begin{document}

\title{Enhancement of the superconducting transition temperature
from the competition between electron-electron correlations and
electron-phonon interactions}
\author{Rayda Gammag$^{1}$ and Ki-Seok Kim$^{2,3}$}
\affiliation{ $^{1}$Asia Pacific Center for Theoretical Physics,
Hogil Kim Memorial building 5th floor, POSTECH, Hyoja-dong, Namgu,
Pohang, Gyeongbuk 790-784, Korea \\ $^{2}$Department of Physics,
POSTECH, Hyoja-dong, Namgu, Pohang, Gyeongbuk 790-784, Korea \\
$^{3}$Institute of Edge of Theoretical Science (IES), Hogil Kim
Memorial building 5th floor, POSTECH, Hyoja-dong, Namgu, Pohang,
Gyeongbuk 790-784, Korea }
\date{\today}

\begin{abstract}
We uncover that the competition between electron-electron
correlations and electron-phonon interactions gives rise to
unexpectedly huge enhancement of the superconducting transition
temperature, several hundreds percent larger ($\geq$ 200 K) than
that of the case when only one of the two is taken into account
($\sim$ 30 K). Our renormalization group analysis claims that this
mechanism for the enhancement of the critical temperature is not
limited on superconductivity but applied to various Fermi surface
instabilities, proposing an underlying universal structure, which
turns out to be essentially identical to that of a recent study
[Phys. Rev. Lett. {\bf 108}, 046601 (2012)] on the enhancement of
the Kondo temperature in the presence of Rashba spin-orbit
interactions. We also discuss the stability of superconductivity
against nonmagnetic randomness.
\end{abstract}


\maketitle

It is our aspiration to increase the superconducting transition
temperature. The BCS (Bardeen-Cooper-Schriefer) theory
\cite{Schriefer_Book} has been our paradigm for the mechanism of
superconductivity. Unfortunately, this fundamental theory does not
allow us to enhance the critical temperature as much as what we
want. Most efforts are dedicated to searching ``the beyond-BCS
theory", where electron correlations are proposed to cause Cooper
pairing, the source of superconductivity. These theories for
unconventional superconductivity can be classified into
Fermi-liquid based theory \cite{Chubukov_Review}, spin-liquid
based theory \cite{Lee_Nagaosa_Wen_Review,Kivelson_Review},
quantum-critical-metal based theory \cite{Varma_Review}, and etc
\cite{Zhang_Review}, based on the mother state for the
superconductivity. Although these theoretical frameworks have
their own predictions in physical spectra, thermodynamics, and
transport for various unconventional superconductors such as high
$T_{c}$ cuprates, organic materials, heavy-fermion systems,
pnictide superconductors, and etc., many fundamental questions
remain unanswered, in particular, even the simplest question: how
can we enhance the critical temperature?

In this letter we revisit this simple but fundamental question,
resorting to the Fermi-liquid based theory and the BCS mechanism,
which allows us to avoid any artificial complexity and give a
definite answer. We introduce both electron-electron correlations
and electron-phonon interactions, which favor $s^{+-}$ and
$s^{++}$ pairing symmetries, respectively, in the $FeAs$-type
two-band structure \cite{FeAs_Review}. The naive expectation is
that the critical temperature ($T_{c}$) decreases down to zero
around the region where both effective interactions become
identical. However, we observe unexpectedly huge enhancement of
$T_{c}$, several hundreds percent larger ($\geq$ 200 K) than that
of the case when only one of the two is taken into account ($\sim$
30 K). Solving coupled BCS gap equations both numerically and
analytically, we prove this $T_{c}$ enhancement.

An interesting aspect of our study is to claim that there exists
an underlying universal structure for the $T_{c}$ enhancement. The
renormalization group analysis clarifies such a structure, which
occurs from Fermi surface instabilities. We observe that our
coupled renormalization group equations for two competing
superconducting correlations are essentially identical to those of
a recent study \cite{Kondo_DM_RG} on the enhancement of the Kondo
temperature, where the interplay between Rashba spin-orbit
interactions and the Kondo effect strengthens the Kondo effect. It
is rather unexpected that both systems have basically the same
structure in the renormalization group sense.

Our model consists of two types of Fermi surfaces, the hole Fermi
surface near the $\Gamma$ point ($c_{\boldsymbol{k}\alpha}$) and
the electron one near the $M$ point ($f_{\boldsymbol{k}\alpha}$),
sometimes regarded as an effective Hamiltonian for $FeAs$
superconductors \cite{FeAs_Review}. We consider two competing
interactions for superconductivity, one of which results from
electron-electron correlations to describe pair hopping from the
hole Fermi surface to the electron one and vice versa,
$H_{\Delta}^{el-el} = \frac{1}{2}
\sum_{\boldsymbol{k},\boldsymbol{p}}
V^{el-el}_{\alpha\beta\beta'\alpha'}(\boldsymbol{k},\boldsymbol{p})
(c_{\boldsymbol{k}\alpha}^{\dagger}c_{-\boldsymbol{k}\beta}^{\dagger}
f_{-\boldsymbol{p}\beta'}f_{\boldsymbol{p}\alpha'} +
f_{\boldsymbol{k}\alpha}^{\dagger}f_{-\boldsymbol{k}\beta}^{\dagger}
c_{-\boldsymbol{p}\beta'}c_{\boldsymbol{p}\alpha'})$, and the
other of which originates from electron-phonon interactions to
introduce conventional BCS pairing on each Fermi surface,
$H_{\Delta}^{el-ph} = - \frac{1}{2}
\sum_{\boldsymbol{k},\boldsymbol{p}}
V^{el-ph}_{\alpha\beta\beta'\alpha'}(\boldsymbol{k},\boldsymbol{p})
(c_{\boldsymbol{k}\alpha}^{\dagger}c_{-\boldsymbol{k}\beta}^{\dagger}
c_{-\boldsymbol{p}\beta'}c_{\boldsymbol{p}\alpha'} +
f_{\boldsymbol{k}\alpha}^{\dagger}f_{-\boldsymbol{k}\beta}^{\dagger}
f_{-\boldsymbol{p}\beta'}f_{\boldsymbol{p}\alpha'})$, where
$V^{el-el(ph)}_{\alpha\beta\beta'\alpha'}(\boldsymbol{k},\boldsymbol{p})
= V^{el-el(ph)}_{\boldsymbol{k},\boldsymbol{p}}
(i\boldsymbol{\sigma}^{y})_{\alpha\beta}
(i\boldsymbol{\sigma}^{y})^{\dagger}_{\beta'\alpha'}$ is the
effective coupling constant for the singlet superconductivity.
Here, we neglect their momentum dependencies, i.e.,
$V^{el-el(ph)}_{\boldsymbol{k},\boldsymbol{p}} = V_{ee(ep)}$.

One may be concerned with spin-density-wave instability, expected
to compete with superconducting instability. Recently, one of us
discussed that marginal breakdown of the Fermi-surface nesting
favors superconductivity instead of spin-density-wave ordering
\cite{FeAs_SC_DMFT}. In this study we assume the regime to favor
superconducting instability instead of spin-density-wave ordering.
We do not take into account the competition between
antiferromagnetism and superconductivity.

We perform the standard mean-field analysis within the BCS
framework. Electron-electron correlations give rise to $s^{+-}$
pairing while electron-phonon interactions result in $s^{++}$
pairing \cite{FeAs_Review}. Introducing this pairing symmetry into
the mean-field analysis, we obtain coupled gap equations,
\begin{widetext}
\bqa && \sum_{\boldsymbol{k}}
\frac{\Delta_{c}}{E_{c}(\boldsymbol{k})} \tanh \Bigl( \frac{\beta
E_{c}(\boldsymbol{k})}{2} \Bigr) + \frac{2V_{ee}}{V_{ee}^{2} -
V_{ep}^{2}} \Delta_{f} + \frac{2 V_{ep}}{V_{ee}^{2} - V_{ep}^{2}}
\Delta_{c} = 0 , \nn && \sum_{\boldsymbol{k}}
\frac{\Delta_{f}}{E_{f}(\boldsymbol{k})} \tanh \Bigl( \frac{\beta
E_{f}(\boldsymbol{k})}{2} \Bigr) + \frac{2V_{ee}}{V_{ee}^{2} -
V_{ep}^{2}} \Delta_{c} + \frac{2 V_{ep}}{V_{ee}^{2} - V_{ep}^{2}}
\Delta_{f} = 0 , \eqa
\end{widetext}
where $\Delta_{c(f)}$ is the total pairing amplitude for the hole
(electron) Fermi surface, which results from both effective
interactions, and $E_{c(f)}(\boldsymbol{k})$ is the conventional
BCS-quasiparticle spectrum \cite{Schriefer_Book} for each Fermi
surface. An interesting point in these equations is the presence
of divergence when both interactions become identical, i.e.,
$V_{ee} = V_{ep}$, implying that the pairing amplitude vanishes.
However, we find a non-monotonic behavior for the critical
temperature near the point of $V_{ee} = V_{ep}$.

\begin{figure}[t]
\includegraphics[width=0.50\textwidth]{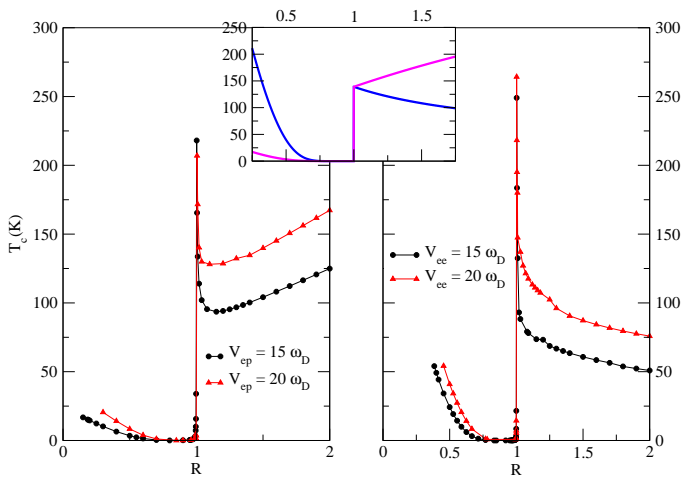}
\caption{Left (Right) : The critical temperature $T_{c}(R)$ from
Eq. (1) as a function of the interaction ratio $R = V_{ee} /
V_{ep}$ with a fixed $V_{ep}$ ($V_{ee}$). The black-circle line is
obtained when $V_{ep} = 15 \omega_{D}$ ($V_{ee} = 15 \omega_{D}$)
and the red-triangle line, $V_{ep} = 20 \omega_{D}$ ($V_{ee} = 20
\omega_{D}$), where $\omega_{D}$ is the Debye frequency. The main
feature is that the critical temperature becomes enhanced at
$V_{ee} \approx V_{ep}$, i.e., $T_{c}(R \approx 1^{+}) \geq 200 K$
($T_{c}(R \approx 1^{+}) \geq 250 K$), several hundreds percent
more than the electron-phonon (electron-electron) driven
critical-temperature $T_{c}(R \rightarrow 0) \approx 25 K$
($T_{c}(R \rightarrow \infty) \approx 30 K$). Inset : The critical
temperature $T_{c}(R)$ from the analytic formula of Eq. (2).
Although the qualitative consistency confirms our numerical
analysis, the quantitative difference results from the choice of
cutoff used to determine $T_{c}(R)$. } \label{fig1}
\end{figure}

Fig. 1 shows how the critical temperature $T_{c}(R)$ changes as a
function of the interaction ratio $R = V_{ee} / V_{ep}$. First, we
start from the regime where the electron-electron interaction is
smaller than the electron-phonon interaction, i.e., $R < 1$ (Left
in Fig. 1). We note that the critical temperature at $R = 0$,
i.e., the electron-phonon driven BCS transition temperature is
about the order of 25 K.
%
%
Increasing electron-electron correlations in $R < 1$, the critical
temperature decreases down monotonically and touches the zero
point before $R = 1$. This result is physically natural, where the
$s^{++}$ pairing competes with the $s^{+-}$ pairing and the total
Cooper-pairing amplitude vanishes near $R = 1^{-}$. On the other
hand, if $R$ increases further to cross the $R = 1$ point, we
observe an abrupt enhancement of $T_{c}$, i.e., $T_{c}(R = 1^{+})
\gg T_{c}(R = 0)$. Further increase of electron-electron
correlations leads $T_{c}(R \gg 1)$ to follow $V_{ee}$ in the
large-$V_{ee}$ region. Second, we start from the regime of $R
\rightarrow \infty$, i.e., $V_{ep} = 0$, where the
electron-correlation driven $s^{+-}$ superconducting transition
temperature is about 30 K (Right in Fig. 1). Increasing the
electron-phonon interaction, the critical temperature also
increases monotonically, touching the highest temperature larger
than 200 K near $R \approx 1^{+}$. $T_{c}$ becomes enhanced
several hundreds percent more! On the other hand, the critical
temperature vanishes near $R = 1^{-}$ when $R$ crosses the $R = 1$
point from the $R > 1$ side. It increases monotonically as $R$
decreases further in $R < 1$, and reaches its saturate value at $R
= 0$ ($V_{ph} \rightarrow \infty$).

One can obtain an analytic expression for the critical temperature
in the case of twin Fermi-surfaces, given by \bqa && T_{c}(R) = D
\exp\Bigl( \frac{1}{N_{F} V_{ee}} \frac{R}{R^{2} - 1} -
\frac{1}{N_{F} V_{ee}} \frac{R^{2}}{|R^{2} - 1|} \Bigr) , \eqa
where $N_{F}$ is the density of states at the Fermi energy and $D$
is the half bandwidth. As shown in Fig. 1 (Inset), this analytic
expression displays qualitatively the same behavior as the
numerical result of $T_{c}(R)$. The quantitative difference
results from the choice of cutoff, used to determine $T_{c}(R)$.

\begin{figure}[t]
\includegraphics[width=0.50\textwidth]{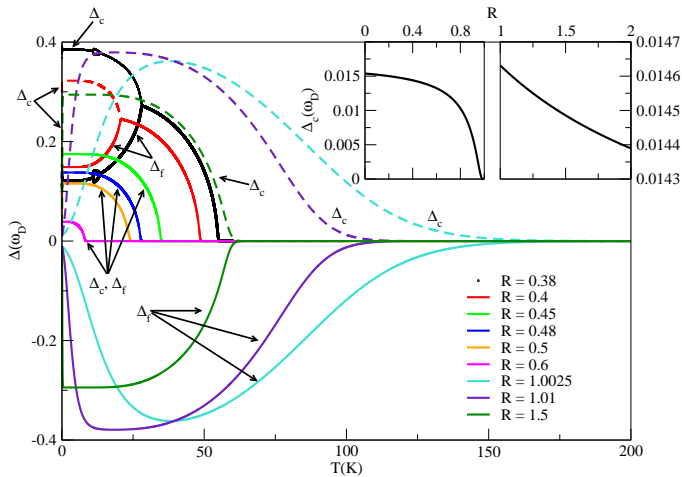}
\caption{Temperature dependencies of both superconducting order
parameters with a fixed $V_{ee}$, where $\Delta_{c(f)}(T)$ is
represented by the dashed (thick) line in $1 \leq R$ while they
are expressed by the upper (lower) curve in $R \ll 1/2$. The main
feature is that both order parameters display non-monotonic
behaviors near $R \approx 1^{+}$, where they become suppressed at
low temperatures, while they exhibit conventional BCS-type
behaviors in both $1/2 \leq R < 1$ and $R \gg 1$. On the other
hand, bifurcation behaviors are observed at low temperatures in $R
\ll 1/2$. We interpret that this bifurcation behavior originates
from the second critical temperature, attributed to the
electron-correlation $s^{+-}$ superconductivity, where
$\Delta_{c(f)}(T)$ increases (decreases) abruptly. Inset : The
total Cooper pairing amplitude $\Delta_{c}(R)$ in the zero
temperature limit, based on the analytic formulae Eq. (3).}
\label{fig2}
\end{figure}


Fig. 2 shows the temperature dependencies of two superconducting
order parameters at each value of $R$. In $1/2 \leq R < 1$ both
$\Delta_{c}(T)$ and $\Delta_{f}(T)$ follow that of the
electron-phonon mediated $s^{++}$ superconductivity while
%
%
the sign difference of $\Delta_{c}(T) > 0$ and $\Delta_{f}(T) < 0$
reflects the $s^{+-}$ pairing symmetry due to electron-electron
correlations in $R > 1$.
%
%
On the other hand, completely unexpected behaviors appear near $R
\approx 1^{+}$. Both superconducting order parameters exhibit
non-monotonic temperature dependencies, where their amplitudes
become suppressed at low temperatures although they start to arise
at much higher temperature than the case of either $R < 1$ or $R
\gg 1$.
%
%
In $R \ll 1/2$ more exotic behaviors are observed. Fig. 2 displays
bifurcation behaviors for order parameters at low temperatures. We
interpret that this bifurcation behavior originates from the
second critical temperature, attributed to the
electron-correlation $s^{+-}$ superconductivity, where
$\Delta_{c(f)}(T)$ increases (decreases) abruptly. However, we do
not understand the reason why this behavior does appear only in $R
\ll 1/2$, not in $R \geq 1/2$.

One can find an analytic formula for the amplitude of the
superconducting order parameter in the zero temperature limit,
\bqa && \Delta_{c}(R < 1; T \rightarrow 0) = D \exp\Bigl( -
\frac{1}{N_{F} V_{ee}} \frac{R}{1 - R} \Bigr) , \nn &&
\Delta_{c}(R > 1; T \rightarrow 0) = D \exp\Bigl( - \frac{1}{N_{F}
V_{ee}} \frac{R}{1 + R} \Bigr) . \eqa Fig. 2 (Inset) shows that
the order parameter at $T = 0$ decreases down to zero
monotonically in $R < 1$ while it becomes almost constant in $1 <
R \leq 2$. Based on this small-amplitude Cooper pairing, we
speculate emergence of anomalous transport phenomena and various
vortex phases under magnetic fields at low temperatures near $R
\approx 1^{+}$ because vortices can be easily created due to the
smallness of the superconducting order parameter and the interplay
between unpaired electrons and vortices will give rise to
anomalies for transport phenomena \cite{Vortex_Review}.

Although the enhancement of $T_{c}$ has been proved within the
mean-field theory, the physical mechanism for the enhancement of
$T_{c}$ has not been identified yet. Performing the standard
renormalization group analysis in the one-loop level, we obtain
coupled renormalization group equations for both effective
interactions, \bqa && \frac{d v_{ep}}{d \ln \mu} = - v_{ep}^{2} -
v_{ee}^{2} , ~~~~~ \frac{d v_{ee}}{d \ln \mu} = - c v_{ep} v_{ee}
, \eqa where both coupling constants are scaled with the density
of states, respectively, and $\mu$ is the typical scale for the
renormalization group analysis. Here, a positive numerical
constant $c$ counts the difference of two Fermi surfaces, and $c =
2$ describes twin Fermi surfaces.

Let's focus on the case of twin Fermi surfaces for simplicity.
Then, these renormalization group equations can be rewritten as
follows, $\frac{d v_{\pm}}{d \ln \mu} = \mp v_{\pm}^{2}$, where
$v_{\pm} = v_{ee} \pm v_{ep}$ are effective coupling constants for
superconductivity. It is straightforward to solve these equations
and find $v_{\pm}(T) = \frac{v_{D}^{\pm}}{1 \mp v_{D}^{\pm} \ln
(D/T)}$ as a function of temperature. The critical temperature is
identified as $v_{\pm}(T_{c}) \rightarrow \infty$, given by $T_{c}
= D \exp\Bigl( - \frac{1}{v_{ee} + v_{ep}} \Bigr)$ when $R
> 1$ and $T_{c} = D \exp\Bigl( \frac{1}{v_{ee} - v_{ep}} \Bigr)$ when $R
< 1$. The second expression results in $T_{c} \rightarrow 0$ as
$v_{ee} \rightarrow v_{ep}^{-}$ in $R < 1$. The first expression
gives rise to the maximum critical temperature when $v_{ee}
\rightarrow v_{ep}^{+}$ in $R > 1$. In fact, these expressions for
$T_{c}$ coincide with the above analytic formula [Eq. (2)] from
the coupled BCS gap equations [Eq. (1)], which crosschecks the
whole procedure of our analysis.

It is rather remarkable to observe that our renormalization group
equations [Eq. (4)] are essentially identical to those in a recent
study \cite{Kondo_DM_RG} on the enhancement of the Kondo
temperature due to the interplay between Rashba spin-orbit
interactions and the Kondo effect. Indeed, one can identify the
electron-phonon interaction with the Kondo coupling constant while
one may match the electron-electron correlation with the Rashba
spin-orbit interaction. In this respect we propose a general
scheme on the way how to enhance the critical temperature, which
occurs from the Fermi-surface instability. The presence of
competing interactions gives rise to the enhancement of the
critical temperature, where ``competition" means that the critical
temperature vanishes or becomes suppressed when one interaction
parameter approaches the other in one direction. More precisely,
the system described by the renormalization group equations, Eq.
(4) is the rule-model for the enhancement of the critical
temperature.

It is important to check the stability of the superconducting
state against weak nonmagnetic randomness since we are considering
unconventional superconductivity and the Anderson theorem
\cite{Anderson_Theorem} does not work in this situation. There are
two types of nonmagnetic impurities. One causes intra-scattering
within the same Fermi surface, and the other generates
inter-scattering between different Fermi surfaces. It has been
both intensively and extensively investigated that the
intra-scattering events do not affect the $s^{+-}$ superconducting
properties such as the critical temperature and gap size much
while the inter-scattering reduce both the critical temperature
and gap size seriously \cite{FeAs_Disorder}. Let's apply these
results to our mixed superconductivity.
%
%
In $R < 1$ the superconducting transition is driven by the
electron-phonon interaction, although the critical temperature
decreases monotonically down to zero as a function of $R$ due to
the renormalization effect from the electron-electron correlation.
It is natural to expect that the Anderson theorem will work in $R
< 1$. However, we can also observe the second critical temperature
in $R \ll 1/2$, given by the electron-electron correlation. As a
result, the low temperature superconductivity will be suppressed
by the inter-scattering events and thus, the abrupt increase or
decrease in superconducting order parameters may appear at much
lower temperature. Furthermore, it is difficult to guarantee that
the Anderson theorem holds in $1/2 \leq R < 1$ because the
amplitude of the superconducting order parameter can be smaller
than the disorder strength. In this respect the competition
between weak $s^{++}$ superconductivity and weak nonmagnetic
randomness will be more complicated. In $R > 1$ the
inter-scattering events will reduce the critical temperature
pretty much. Short-range nonmagnetic impurities will give negative
effects on the enhancement of $T_{c}$. More quantitative and
self-consistent analysis is necessary.

In conclusion, we proposed a general scheme on the way how to
enhance the critical temperature for superconductivity. An
essential ingredient is to introduce competing interactions. Here,
we considered both electron-phonon interactions and
electron-electron correlations and demonstrated that the
superconducting transition temperature becomes much enhanced more
than several hundreds percent when the electron-phonon interaction
increases up to the electron-electron correlation. We claim that
this phenomenon will not be limited on superconductivity but be
generalized to various Fermi-surface instabilities. The
renormalization group analysis uncovers the underlying general
structure for the enhancement of the critical temperature, given
by Fermi-surface instabilities. Indeed, the Kondo effect in the
presence of the Rashba spin-orbit interaction \cite{Kondo_DM_RG}
shows essentially the same renormalization group structure as the
electron-electron vs. electron-phonon competing superconductivity.

KS would like to thank P. Fulde for introducing us this problem.
KS also thanks T. Takimoto, S. P. Mukherjee, M.-T. Tran, and J.-H.
Han for helpful discussions. KS was supported by the National
Research Foundation of Korea (NRF) grant funded by the Korea
government (MEST) (No. 2012000550).

\end{document}